\documentclass[12pt,aps,prb,reprint,showpacs]{revtex4-1}   

\usepackage{amsmath}    
\usepackage{graphicx}   
\usepackage{multirow}
\usepackage{rotating}
\usepackage{lipsum}
\usepackage{hyperref}
 
\begin{document}
\title{Developing Peer Instruction questions for quantitative problems for an upper-division astronomy course}

\author{Colin S.\ Wallace}
\affiliation{Department of Physics and Astronomy, University of North Carolina at Chapel Hill, Chapel Hill, North Carolina 27599} 
\email{cswallace@email.unc.edu} 

\date{\today}

\begin{abstract}

Decades of research show that students learn more in classes that utilize active learning than they do in traditional, lecture-only classes.  Active learning also reduces
the achievement gaps that are often present between various demographic groups.  Given these well-established results, instructors of upper-division astronomy courses 
may decide to search the astronomy education research literature in hopes of finding some guidance on common student difficulties, as well as research-validated and 
research-based active learning curricula.  Their search will be in vain.  The current literature on upper-division astronomy is essentially non-existent.  This is a shame,
since many upper-division astronomy students will experience conceptual and problem-solving difficulties with the quantitative problems they encounter.  These difficulties 
may exist even if students have a strong background in mathematics.  
In this paper, I examine one quantitative problem that is representative
of those that upper-division astronomy students are expected to solve.  I list many of the subtle pieces of information that students need to understand in order to 
advance toward a solution and I describe how such a list can be used to generate Peer Instruction (PI) questions.  
I also provide guidelines for instructors who wish to develop and implement their own PI questions.
These PI questions can be used to increase the
amount of active learning that occurs in an upper-division astronomy course.  They help develop students' understandings of symbolic, mathematical representations and they 
help improve students' problem-solving skills.  
The ideas presented in this paper can help instructors infuse
their upper-division astronomy courses with active learning.

\end{abstract}

\maketitle

\section{Introduction}
\label{intro}

An instructor teaching an upper-division astronomy course faces many challenges.  They must decide what exactly they will teach; unlike physics, there is no widely accepted 
``canon" of topics.  There is not even a
consensus about which courses astronomy majors should take and in what order.  Across undergraduate astronomy degree programs, there is more uniformity in the physics 
course requirements than in the astronomy course requirements.  The only courses that are required for all astronomy bachelor degrees, regardless of institution, are 
introductory calculus-based mechanics and electricity and magnetism, as well as the necessary calculus pre-requisites.\cite{cabanela03}  

Astronomy courses are also frequently organized
around specific objects to be studied (e.g., stellar physics, the interstellar medium, cosmology, etc.), whereas physics courses are organized according to physical principles 
(e.g., classical mechanics, quantum mechanics, thermodynamics and statistical physics, etc.).  Principles from across the physics curriculum often appear in a single 
upper-division astronomy course.  For example, a complete understanding of stellar physics requires ideas drawn from classical mechanics, electricity and magnetism, quantum
mechanics, thermodynamics and statistical physics, nuclear physics, and relativity.  Upper-division astronomy instructors cannot assume that their students have completed 
upper-division physics coursework in all of these sub-disciplines.  Astronomy instructors must be prepared to provide their students with their first lessons on these 
topics at
an advanced undergraduate level.  This means that astronomy instructors must also be able to help their students overcome common and well-documented difficulties in 
upper-division physics while simultaneously teaching them astrophysics.  Students are likely to experience conceptual, reasoning, and problem-solving difficulties that
are specific to astrophysics topics, but unfortunately the community's knowledge of what these difficulties might be is limited to a single paper that found that introductory 
astronomy students experience the same conceptual difficulties regardless of whether they are in a course designed for STEM or non-STEM majors. \cite{zeilik03}

On top of all of these challenges, upper-division astronomy instructors also have an obligation to use active learning techniques to intellectually engage their students.
No one can reasonably doubt the benefits of active learning.  Decades of research show that active learning increases the amount of information that students learn and 
retain.\cite{freeman14}  Active learning also reduces pernicious achievement gaps between different demographic groups -- gaps that are exacerbated by traditional, 
lecture-only instruction.\cite{black98,eddy14,lorenzo06,rudolph10}  Instructors must incorporate active learning into their courses if they truly care about creating inclusive and 
effective learning environments.  This 
presents another challenge for upper-division astronomy instructors who wish to use research-based active learning curricula.  When they consult the literature, they will find that
the astronomy education community (including this author) have focused almost all of their efforts at the college level on Astro 101 courses. \cite{bailey18,bailey05,lelliott10}  
This presents a sad contrast to our colleagues in physics, 
who have created numerous active learning materials, based on extensive physics education research, for multiple upper-division physics courses.\cite{loverude15}  Hopefully, the 
astronomy community will someday catch up with our peers in physics, but in the meantime what are astronomy instructors supposed to do?

In this paper, I describe how instructors of upper-division astronomy courses can take a standard quantitative problem (similar to those that students are expected to solve 
on homework and exams) and develop a series of Peer Instruction (PI) questions that help students unpack the essential steps toward a 
solution and decode the oft-subtle symbolic representations used by professionals.\cite{mazur97}  Note that PI questions are often referred to as Think-Pair-Share (TPS) in the 
astronomy community, in recognition of the many similarities between Mazur's Peer Instruction and Lyman's original conceptualization of TPS.\cite{lyman81}

I am focusing on PI questions for upper-division courses for two reasons.  First, PI is the active learning technique that is most frequently adopted by instructors.\cite{dancy10}
Second, the transition from the introductory to upper-division courses is known to be difficult.
\cite{manogue06}
Even students who excelled in their introductory courses may struggle once they advance into upper-division courses sometime around their sophomore or junior years.
Many instructors recognize that these struggles are tied to the increased use of mathematics.\cite{wilcox13}  Instructors often lament the mathematical preparation of
some students, and it is absolutely true that an inadequate foundation in mathematics will severely hamper a student's ability to master the material presented at the 
upper division.  However, the ways in which mathematics is presented and used in math classes does not always match how it is presented and used in physics and astronomy,
so simply encouraging students to ``take more math" or to ``do better in math" will not necessarily fix all of the issues they experience when attempting to solve quantitative
problems.\cite{dray99, redish05}  For example, students may understand how to do a triple intergral and yet struggle to understand what certain symbols represent, why certain substitutions are valid and essential, and 
how we use our knowledge to set limits and make reasonable assumptions that simplify the problem.  

At this time, it is necessary to use an example to illustrate this point.  In Section \ref{pressure}, I present a typical calculation that many astronomy majors will encounter and I 
highlight 
the various steps that are not clearly explicated in standard solutions found in a wide variety of textbooks.  In Section \ref{PI}, I show how these steps can be transformed into
PI questions that can be used to actively engage students, with the concomitant effect of demystifying many of the intermediate steps highlighted in Section \ref{pressure}.  
In Section \ref{dev_imp} I present guidelines for instructors who wish to develop and implement their own PI questions.  Conclusions can be found in Section \ref{conc}.

\section{The Pressure Integral}
\label{pressure}
Here is a problem that many astronomy majors may encounter at some point during their undergraduate studies:
\begin{quote}
Imagine that an interstellar dust grain is surrounded by an isotropic gas.  $n(\theta,\phi,v)$ represents the number density of gas particles with speeds between $v$ and 
$v + dv$ coming from polar angles between $\theta$ and $\theta + d\theta$ and azimuthal angles between $\phi$ and $\phi + d\phi$.  Each particle in the gas has the same 
mass $m$.  Find an expression for the pressure $P$ that these gas particles exert on the dust grain.  Assume the dust grains reflect specularly. 
\end{quote}
The first few steps in the solution may look like the following:
\begin{gather*}
P = \int_{0}^{2\pi} \int_{0}^{\pi/2} \int_{0}^{\infty} \left( 2mv\ \textnormal{cos}(\theta) \right) \left( n(\theta,\phi, v) \right) \\
v\ \textnormal{cos}(\theta)\ \textnormal{sin}(\theta)\ dv\ d\theta\ d\phi \\
= \int_{0}^{\infty} 2mv^{2} \left( n(\theta,\phi,v) \right) dv 
= \frac{1}{3} m \int_{0}^{\infty} v^{2} n(v)\ dv \\
\end{gather*}
At this point, we need to specify a distribution function for the number of particles in order to evaluate the final integral, which is often called the pressure integral.\cite{carroll07} 
But for the purposes of this paper, we need not go any further.  There are already several pieces of important information that are embedded in the above 
calculation -- information that may be missed by many students.

Variants of this calculation are found in a number of astrophysics textbooks.\cite{carroll07,harwit06,maoz07}  Different textbooks provide different amounts of exposition
surrounding the details of the calculation.  
Many will invoke impulse-momentum principles to explain why the 
$2mv\ \textnormal{cos}(\theta)$ term is present.
Many will also provide a sentence or two to attempt to explain the origins of the second cosine term.  
Some may also indicate how one goes from an expression involving $n(\theta,\phi, v)$
to one involving $n(v)$.  But a student unaccustomed to the density of information packed into the problem statement and the symbolic notations will require substantially more
scaffolding to link the mathematical expressions to an understanding of the physical scenario being modeled.  

A conscientious instructor may look at the above calculation and ask themselves the following questions:
\begin{itemize}
\item Do my students recognize that the set of terms $n(\theta,\phi,v)\ v$ is proportional to a collision rate?
\item Do they really understand why the collision rate is multiplied by cos($\theta$)?
\item Do my students know that the $2mv\ \textnormal{cos}(\theta)$ represents a change in momentum?
\item Do students understand that a change in momentum is proportional to a pressure?
\item Do students understand why a differential element of solid angle $d\Omega$ is given by sin$(\theta)\ d\theta\ d\phi$?
\item Would my students have chosen the same limits of integration?
\item A factor of $4\pi$ must appear in the denominator when going from $n(\theta,\phi, v)$ to $n(v)$.  Can my students explain why?
\end{itemize}  
This list is not meant to be comprehensive.  But it is extensive enough to make a key point: Struggling on a problem such as this is not a simple matter of being poor at 
math.  A student may have done quite well in calculus and may be able to evaluate a large number of complex integrals.  At first glance, this looks like
yet another triple integral that he or she should be able to do.  However, if that student lacks the symbolic and representational fluency necessary to answer all of the questions 
posed above, then they will be unable to make much progress toward a solution, much to their consternation and the consternation of their instructor.  

\section{Sample PI Questions}
\label{PI}
Instructors should not be discouraged by the questions listed in Section \ref{pressure}.  Instead, they should see each question as an indication of a topic to explore 
in detail using active learning techniques.  While there many such techniques that could be used, in this paper I will focus on Peer Instruction.  PI questions are often the
first technique adopted by instructors who are new to active learning.  Instructors report using this technique more than any other.\cite{dancy10}  PI is
the gateway drug to active learning.

\begin{figure}
\includegraphics[scale=1]{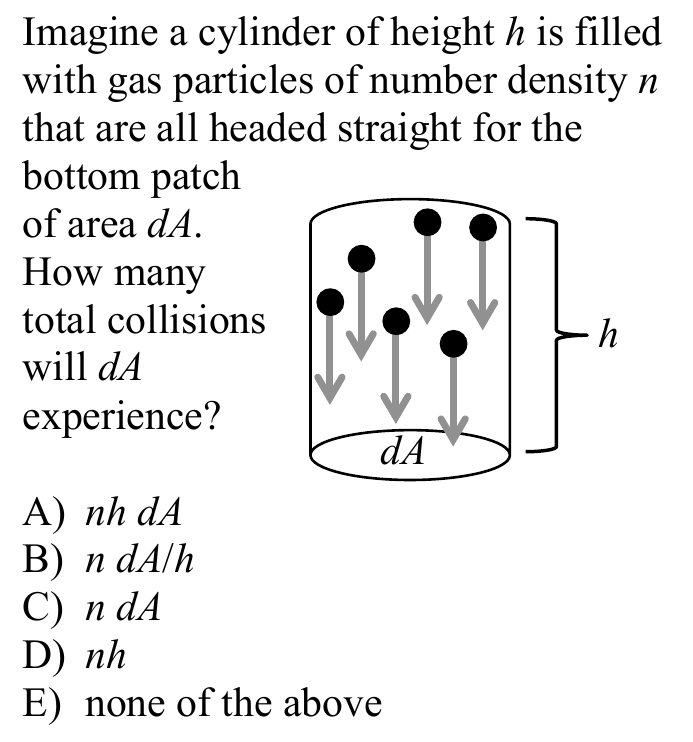}
\caption{\label{figure1}A sample PI questionon how the number of collisions depends on the number density of particles, the height of the cylinder, and the patch of area $dA$.  The correct answer is A.}
\end{figure}

\begin{figure}
\includegraphics[scale=1]{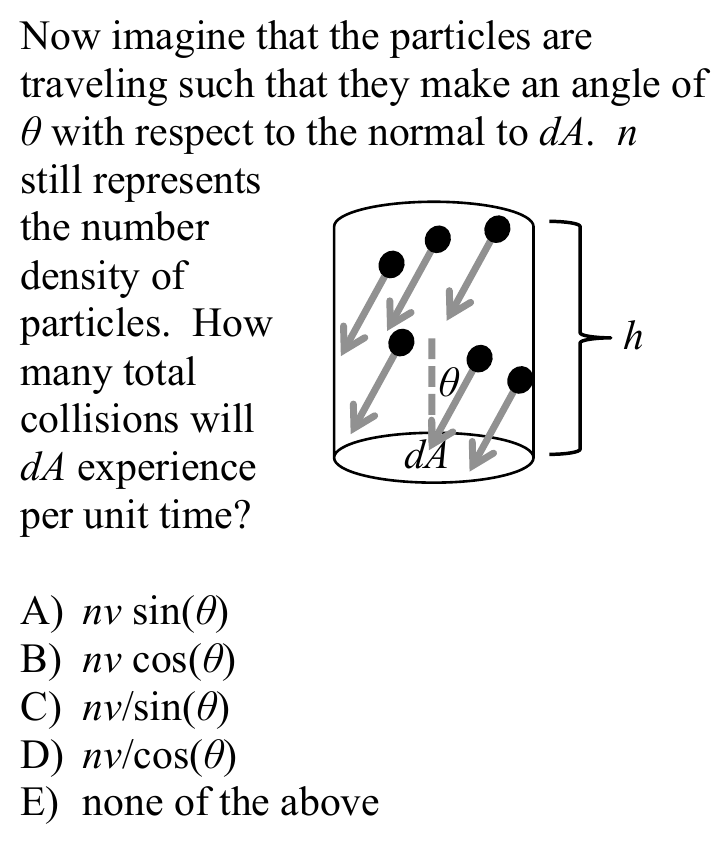}
\caption{\label{figure2}A sample PI question on how the number of collisions per unit time depends on the angle at which the particles strike the surface.  The correct answer is B.}
\end{figure}

\begin{figure}
\includegraphics[scale=1]{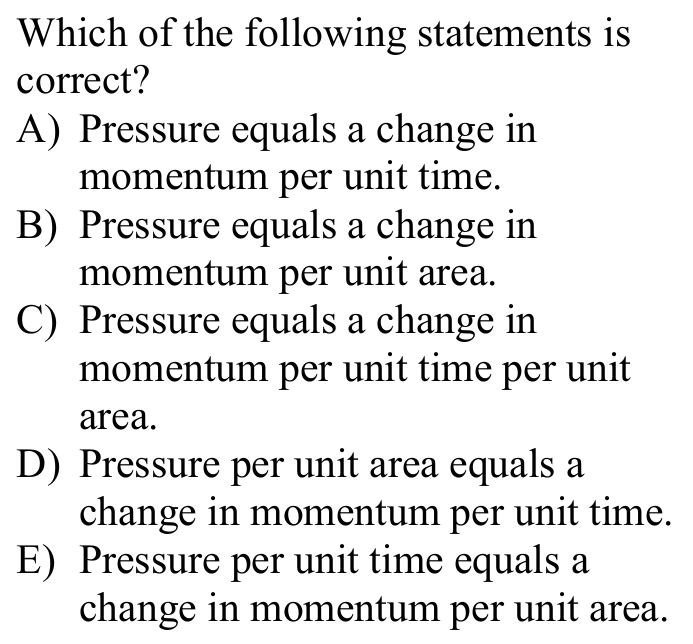}
\caption{\label{figure3}A sample PI question about the relationship between pressure and a change in momentum.  The correct answer is C.}
\end{figure}

\begin{figure}
\includegraphics[scale=1]{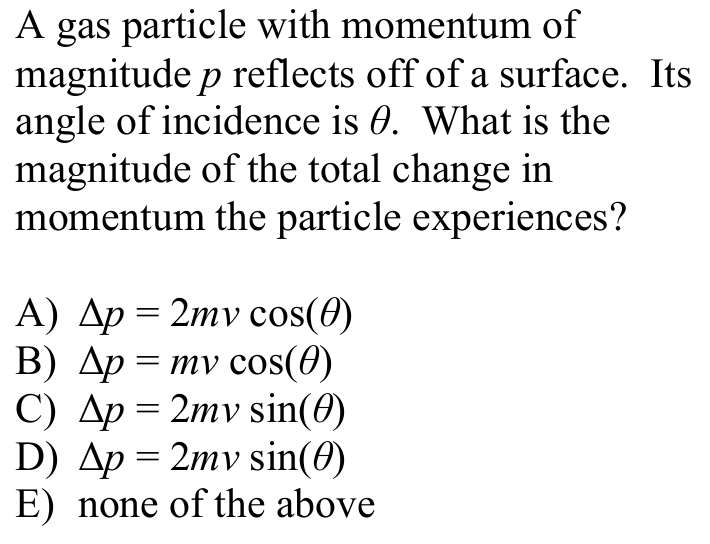}
\caption{\label{figure4}A sample PI question on the change of momentum experienced by a particle colliding with a surface.  The correct answer is A.}
\end{figure}

\begin{figure}
\includegraphics[scale=1]{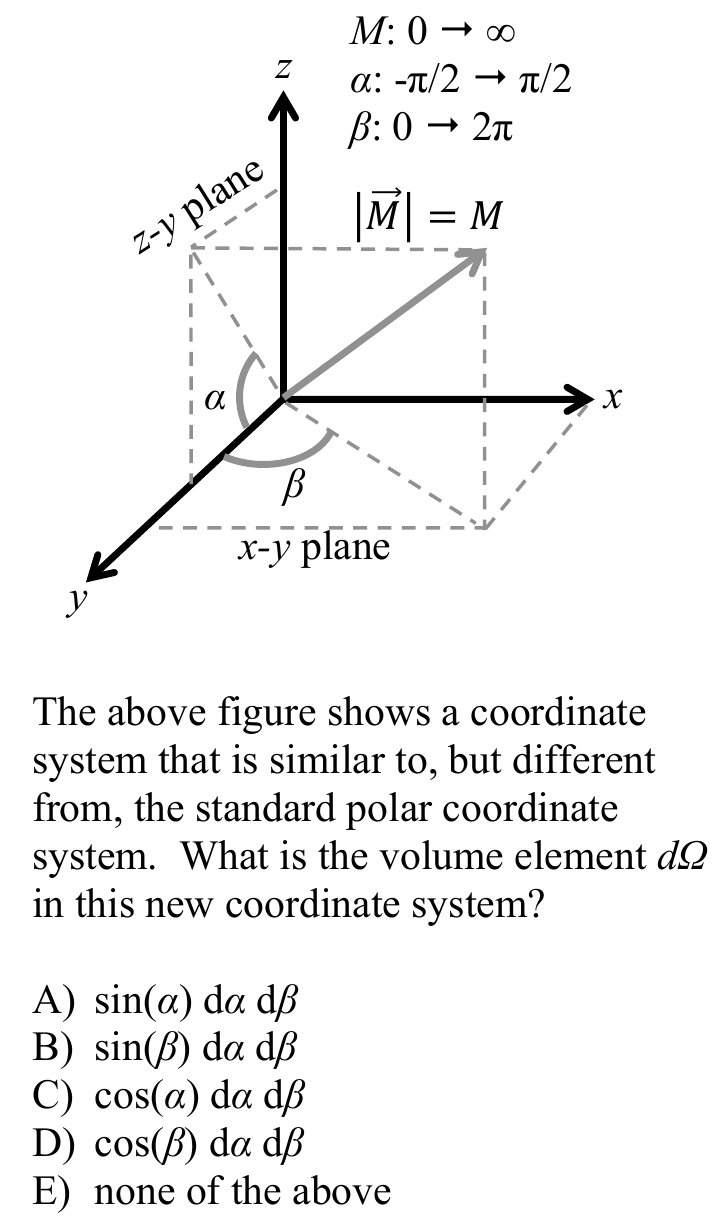}
\caption{\label{figure5}A sample PI question on determining the volume element for a given coordinate system.  The correct answer is C.}
\end{figure}

\begin{figure}
\includegraphics[scale=1]{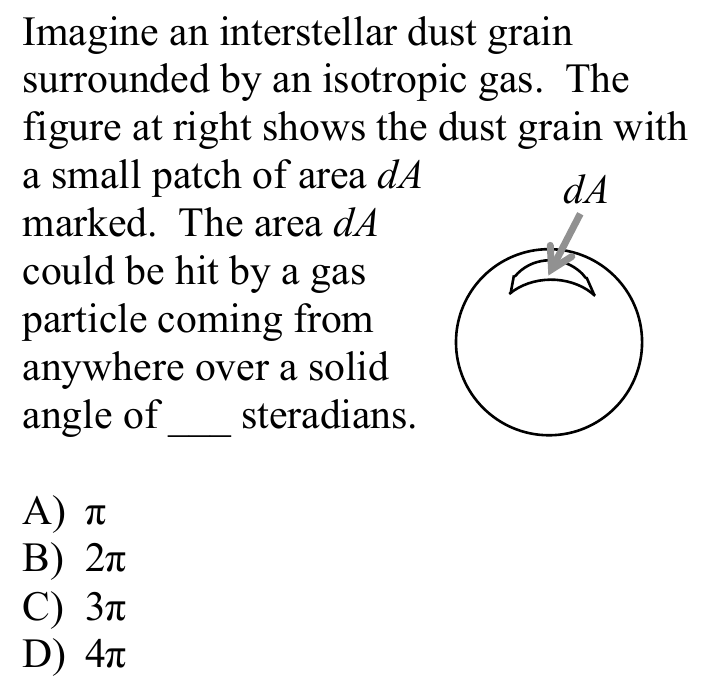}
\caption{\label{figure6}A sample PI question on the solid angle from which particles can originate if they are to collide with a patch of area $dA$.  The correct answer is B.}
\end{figure}

\begin{figure}
\includegraphics[scale=1]{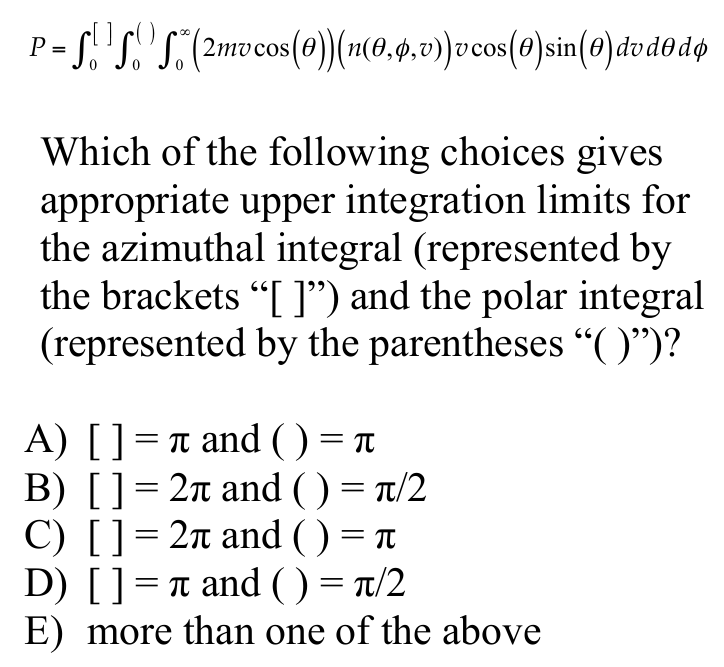}
\caption{\label{figure7}A sample PI question about choosing the appropriate integration limits.  The correct answer is E.}
\end{figure}

\begin{figure}
\includegraphics[scale=1]{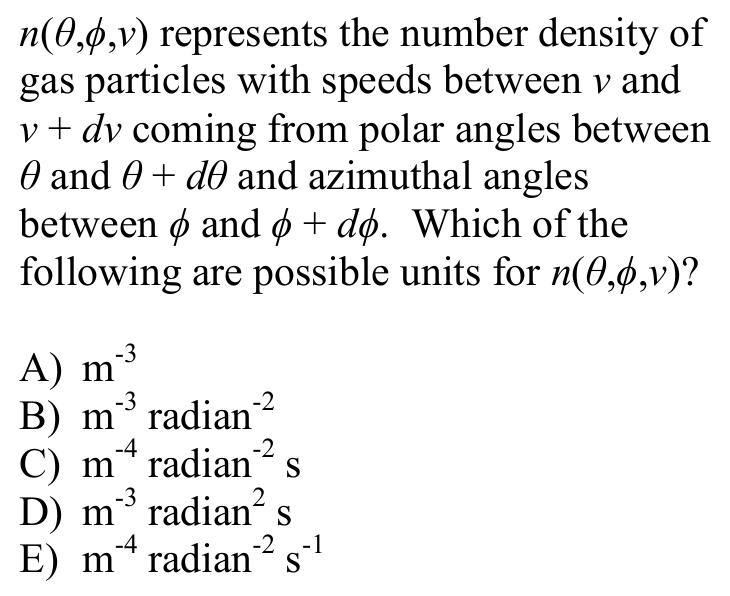}
\caption{\label{figure8}A sample PI question about the appropriate units for $n(\theta,\phi, v)$.  The correct answer is C.}
\end{figure}

\begin{figure}
\includegraphics[scale=1]{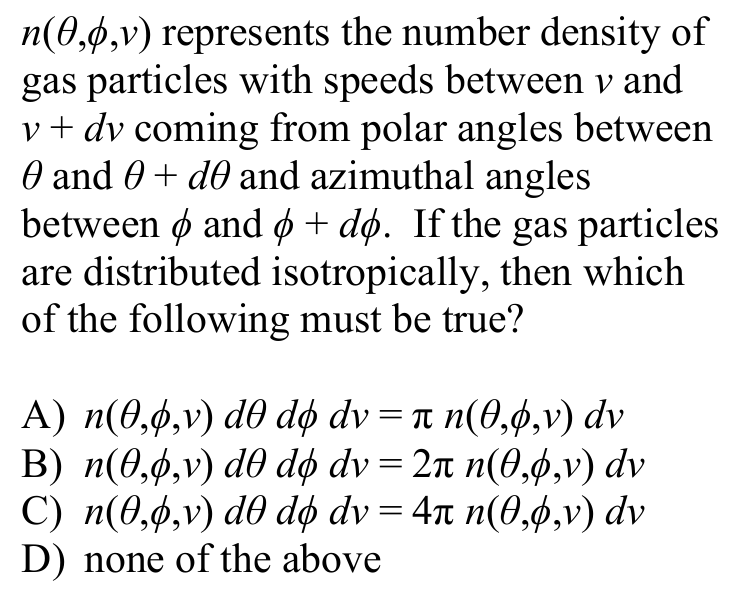}
\caption{\label{figure9}A sample PI question about the correct expression for $n(\theta,\phi, v)\ d\theta\ d\phi\ dv$.  The correct answer is C.}
\end{figure}

\begin{figure}
\includegraphics[scale=1]{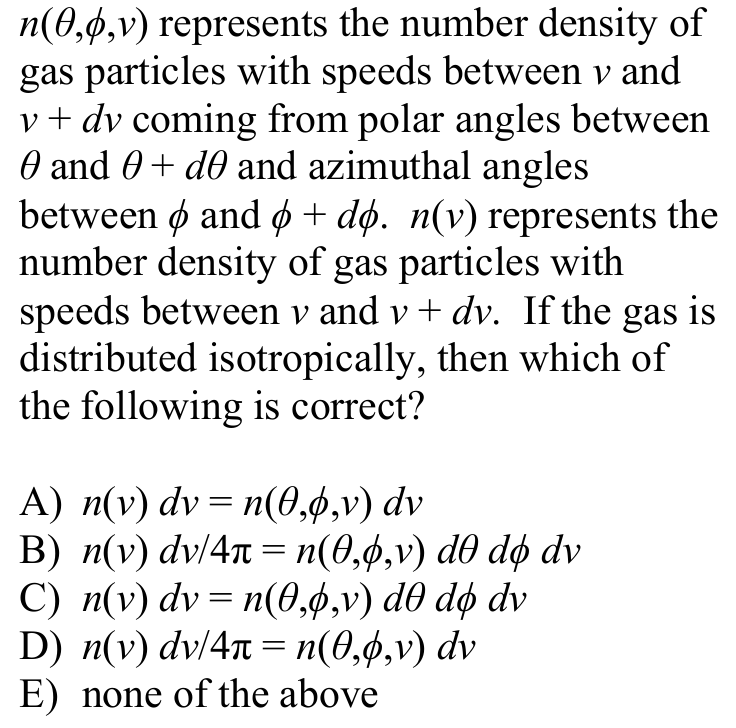}
\caption{\label{figure10}A sample PI question about the correct expression for $n(v)\ dv$ for an isotropic gas.  The correct answer is D.}
\end{figure}

Each of the questions listed above can be turned into one or more PI questions.  Figures \ref{figure1}-\ref{figure10} are a possible sequence of PI
questions that instructors can use to help guide students through the potential difficulties raised in Section \ref{pressure}.  

The first two questions (Figures \ref{figure1} and \ref{figure2}) are designed to help students understand why the number of collisions per unit time per unit area is given by $nv\ \textnormal{cos}(\theta)$.
The PI question in Figure \ref{figure1} presents students with a simplified scenario in which the number of collisions is given by $nh\ dA$.  Instructors can build
on this result to explain that the number of collisions is also equal to $n v \Delta t\ dA$, where $v$ is the particle's speed and $\Delta t$ is the time interval required for particles 
at the top of the cylinder to travel all the way down to the area $dA$, which is why the number of collisions per unit time per unit area for this scenario in Figure
\ref{figure1} is $nv$.  The PI question in Figure \ref{figure2} forces students to think about how to modify the expression for the number of collisions per unit time per unit area to account
for a situation in which the particles' velocity makes an angle of $\theta$ with respect to the normal to $dA$.  This question leads students to understand the origin of one of the
$\textnormal{cos}(\theta)$ terms in the pressure integral.

Students must next connect $nv\ \textnormal{cos}(\theta)$ with the concepts of 
momentum and pressure.  Figures \ref{figure3} and \ref{figure4} address this issue.  The question in Figure \ref{figure3} forces students to make a 
conceptual connection between pressure and a change in momentum.  The PI question in Figure \ref{figure4} requires students to use their knowledge 
of geometry and impulse-momentum principles to determine that the 
there must be a second cos($\theta$) term in the pressure integral.

At this point, students should be ready to write down the triple integral that they will need to evaluate.  Many will recognize that they are integrating over a solid
angle element $d\Omega$.  Some may remember that $d\Omega =$ sin$(\theta)\ d\theta\ d\phi$ for the standard spherical coordinate system used in physics and astronomy.
But do they really understand why $d\Omega =$ sin$(\theta)\ d\theta\ d\phi$?  Recent research shows that the answer for a majority of students is ``no."\cite{scher19}
Figure \ref{figure5} shows a PI question, adapted from the work of Schermerhorn and 
Thompson (2019), that makes students think about how to construct a volume
element for a given coordinate system.\cite{scher19}

Students should also wrestle with determining the limits of integration, especially for the polar and azimuthal integrals.  One cannot simply state that these integrals must be
done over ``all possible angles."  Na\"{i}vely integrating over $\theta$ from 0 to $2\pi$ will result in an answer of 0!  A more careful student may integrate over $\phi$ from
0 to $2\pi$ and $\theta$ from 0 to $\pi$ (or the equivalent of half a circle), but this will result in an answer that is a factor of 2 too big.  Figures \ref{figure6} and \ref{figure7} address
this issue.  Figure \ref{figure6} is designed to help student realize that they only need to integrate over a solid angle of $2\pi$
steradians, while Figure \ref{figure7} is designed to focus students' attention on which integration limits are appropriate.

Finally, a student who simply substitutes $n(v)$ in place of $n(\theta,\phi, v)$ will find that their answer is too big by a factor of $4\pi$.  To understand why, this student must
realize that $n(\theta,\phi, v)$ is the the number density of gas particles with speeds between $v$ and $v + dv$
coming from polar angles between $\theta$ and $\theta + d\theta$ and azimuthal angles between $\phi$ and $\phi + d\phi$.   Similarly, $n(v)$ is the number density of 
gas particles with speeds between $v$ and $v + dv$.  The PI question in Figure 
\ref{figure8} probes 
whether students really understand what $n(\theta,\phi, v)$ represents by asking about its possible units.  An instructor can build off of this 
question by pointing out that the number density of particles is given by $n(\theta,\phi, v)\ d\theta\ d\phi\ dv$.

The PI question shown in Figure \ref{figure9} builds upon this idea.  If the particles are distributed isotropically, then the number density of particles is 
$n(\theta,\phi, v)\ d\theta\ d\phi\ dv = 4\pi n(\theta,\phi, v)\ dv$.  Both $4\pi n(\theta,\phi, v)\ dv$ and $n(v)\ dv$ represent the number density of particles,
so 
\begin{equation*}
n(\theta, \phi, v) dv = \frac{n(v)\ dv}{4\pi} ~.
\end{equation*}
This idea is probed by the PI question in Figure \ref{figure10}.

The series of PI questions depicted in Figures \ref{figure1}-\ref{figure10} represent a possible way to introduce active learning into an upper-division astronomy class.  
While PI questions have traditionally been used to target students' conceptual difficulties, the PI questions shown above specifically address difficulties students may have when solving a
quantitative problem.  Many of the PI questions have answer choices that are involve mathematical expressions, and the incorrect choices use expressions that exemplify common
difficulties that many students experience when they attempt to translate a physical situation into a mathematical representation.   
The PI questions shown in Figures \ref{figure1}-\ref{figure10} are not meant
to be a definitive list of all the PI questions instructors can and should ask of students who are studying the pressure integral, but they do demonstrate
that it is possible to use PI to active engage upper-division astronomy students on quantitative problems.

\section{Developing and Implementing PI Questions}
\label{dev_imp}
Now that I have unpacked the pressure integral problem, listing a series of PI questions to scaffold students through the multiple reasoning steps required to reach a solution, 
I will provide some guidelines for instructors who wish to develop and implement their own PI questions.  These guidelines originate from multiple sources.  Many derive 
from the known best practices for implementing PI.\cite{TPS}  Others come from thoughtful reflection on my own experiences using PI questions that I have created in order
to facilitate collaborative problem-solving during the ``lecture" portion of a class.  Some guidelines emerged from critiques and conversations I have had with colleagues.
Since quantitative problem-solving is a skill taught across the physics and astronomy curriculum (as well as other disciplines), the guidelines listed here apply to a wide variety
of courses, from the introductory to the graduate level.  However, since this paper is focused on upper-division astronomy, I will limit the examples I use to that context.

\subsection{Developing PI Questions}
\label{dev}
If you are an instructor who wants to use PI questions to scaffold the in-class solving of quantitative problems, then you must begin by identifying which problems you want 
your students to solve.  If you have taught the course before, recall which problems were notoriously difficult for your students.  Many of us have had the experience
of delivering what we consider to be lucid and informative lectures, replete with multiple example problems whose solutions we model for students, only to have the demoralizing 
experience of watching our students struggle on homeworks, on exams, and during office hours to apply or even recall the information and procedures explained during lecture.
When students consistently struggle with a particular problem or problem-solving technique, then consider using a sequence of PI questions to help students work toward a solution
during class.

Even if you are teaching your course for the first time, you should be able to identify a set of quantitative problems that require students
to use the range of problem-solving techniques you plan to teach.  While you will want to put some of these problems on homework sets and save others for your exams, you 
should also use some during class.  I recommend using some of your most complex problems during class.  Class time is precious and limited, 
so while it is tempting to try to have your students work through an entire progression of problems, from the relatively straightforward to the increasingly complex,
you may find that you only have time for your students 
to do one or two problems in a single class period -- and if you only have time for one problem, it better engage your students in as many reasoning steps and problem-solving techniques as 
possible.  Problems that require students to only use a limited number of problem-solving techniques may not adequately prepare students
for the more complex problems they will encounter that require coordinating multiple pieces of information.  In contrast, when students attempt to solve a single complex problem during 
class, you can then use a series of PI questions to help direct the class conversation to the exact parts of the problem-solving process that will be the most difficult for your students.

But how can you tell which parts of that process will be the most difficult for students?  If you have taught the course before, then you may already have a sense of where your students
might struggle.  For example, many instructors have
worked with students who struggle to correctly identify and place a trigonometric function in an equation.  A question such as the one shown in Figure \ref{figure2} targets this difficulty since sine and 
cosine appear in the answer choices, both in the numerator and the denominator.  If you have yet to identify
exactly where students may be struggling on a particular problem, then review how you solved the problem.  Every time you performed a step that involved more than plugging
in a number, performing an algebraic manipulation, or executing a well-known algorithm (such as integrating a polynomial) -- in other words, every time you had to really think about what to 
do next -- you have identified a step with which many students will likely struggle.  These are the steps for which you should develop PI questions.

Look again at the series of PI questions described in Section \ref{PI}.  They all target places along the problem-solving pathway that require students to do more than push a few symbols 
around or plug in a number.  The PI questions in Section \ref{PI} ask students to identify integration limits, use a coordinate system to accurately describe positions, construct representations
of physical quantities using mathematical symbols, and use mathematical representations learned in prior courses.  While these are not the only cognitive steps that
students may struggle with, they will encompass a large fraction of students' difficulties, and instructors trying to develop their own PI questions can mimic some of the PI questions shown
in Section \ref{PI}. 

For example, if you want your students to integrate Planck's law in order to derive the Stefan-Boltzmann law, then you will need to make sure that they can 1) understand why a cos($\theta$) term appears 
in the integrand alongside Planck's law, and 2) recognize that they only need to integrate over half a sphere.  Consequently, you may want to adapt the PI
questions shown in Figures \ref{figure2} and \ref{figure7} for this case.  Even though the derivation of the Stefan-Boltzmann law from Planck's law may seem like a purely mathematically exercise at first glance,
incorportaing the cos($\theta$) term and choosing the appropriate integration limits requires students to visualize the physical scenario they are modeling and to understand what the various quantities represent.
From my experience, students are especially likely to struggle with choosing the integration limits; initially,  many want to integrate over all possible polar and azimuthal angles, suggesting that they do not immediately 
understanding the reasoning for integrating over only half of a sphere.

To take a second example, imagine that you have just introduced your class to Einstein's $A$ and $B$-coefficients.  Before proceeding, you may want to see if your students can generate the correct relationship
between these three coefficients (and the density of photons or energy density or mean intensity) under the condition of thermodynamic equilibrium.  You could ask a PI question such as ``Which of the following
expressions correctly relates Einstein's coefficients under the condition of thermodynamic equilibrium?" The correct answer is the mathematical relationship $n_{L} B_{LU} \bar{J}  = n_{U} A_{UL} + n_{U} B_{UL} \bar{J}$,
where $n_{L}$ and $n_{U}$ are the number densities of atoms at the lower and upper energy states, respectively, and $\bar{J}$ is the mean intensity.  The distractors (wrong answers) can include variations 
on this expression that place a coefficient on the wrong side of the equals sign (e.g., $n_{U} A_{UL} = n_{L} B_{LU} \bar{J} + n_{U} B_{UL} \bar{J}$) and/or incorrectly includes or excludes $\bar{J}$ in certain 
parts of the equation (e.g., $n_{L} B_{LU} \bar{J}  = n_{U} A_{UL} \bar{J} + n_{U} B_{UL} \bar{J}$), among other errors.  A PI question like this forces students to practice contructing representations of
astrophysical ideas using mathematical symbols, a skill that students might not have the opportunity to develop during lecture if the instructor simply writes the answer on the board for students to dutifully copy down
into their notebooks.

As a final example, imagine your upper-division class is studying cosmology and you want them to determine the current proper distance between an observer and a distant galaxy in a flat, matter-dominated 
universe described by the Friedmann-Robertson-Walker (FRW) metric.  There are multiple steps along this problem's solution pathway where students must make a decision.  Students can benefit from a PI question at 
each of these steps.  You can ask students whether the $d\Omega^2$ term in the FRW metric should be multiplied by $r^2$, sin$^2 (r)$, or sinh$^2(r)$ for a flat universe.  You can ask whether the scale factor
$a$ scales as the time coordinate $t$ according to $a \sim t^{1/2}$, $a \sim t^{2/3}$, or $a \sim e^{t}$ for a matter-dominated universe.  You can ask students which of the differentials ($ds$, $dt$, $dr$, $d\theta$, and $d\phi$)
can be set equal to 0.  These are some of the major decisions students have to make before they can carry out the algebra and calculus necessary to solve this problem.  Once students
understand the reasoning underlying each of these decisions, they can apply what they have learned to new situations (e.g., calculating the proper distance in an open, radiation-dominated universe).  PI questions can
help you foster the conversations your students need to have during class in order to develop this understanding.

Please note that it is not the purpose of this section to develop a comprehensive list of all possible PI questions for all possible upper-division astronomy problems.  Instead, I am using a few examples to show how one 
can generalize the approach demonstrated in Section \ref{PI} to create PI questions appropriate for scaffolding and developing upper-division astronomy students' quantitative problem-solving abilities.
All of the PI questions suggested in this section and shown in Section \ref{PI} focus students' attention on the key decisions that one must make when solving a particular problem.  When a student struggles
to successfully solve a quantitative problem, they are most likely struggling with one or more of these decisions, which may include choosing integration limits, applying a coordinate system, and constructing 
mathematical representations.  PI questions that target these decision points can stimulate productive student-student and student-instructor conversations that help students better understand the reasoning they
need to employ when solving quantitative problems. 

\subsection{Implementing PI Questions}
\label{imp}
Research suggests that the quality of an instructor's implementation of active learning is the most important factor in determining their students' learning gains.\cite{prather09}
This section describes some of the best practices that my colleagues and I have developed for implementing PI questions, especially in the context of scaffolding the solving of
a quantitative problem.  These implementation guidelines are meant to help you ensure that students are fully engaged with the problem and that you are efficiently using your limited class time.

For some instructors, time is a barrier that discourages them from using an active learning strategy, such as PI.\cite{dancy10}  To be honest, if you attempt to simply squeeze in some active learning on top of the traditional lectures that you have given in the past, then time \emph{will} be an issue.  Both your lectures and your active 
learning activities will be cut short, likely before you get to some of the more interesting -- and complex -- topics.  This will cause both you and your students to be frustrated.  To avoid this frustration, you will need to decide 
how to most efficiently use the available class time, which will require you to make judicious decisions about your lecture topics.  This does not necessarily mean that you have to dump a large fraction of the
content you would otherwise lecture on, although some trimming may be necessary.  Rather, you will have to decide what content students are best equiped to pick up outside of class (e.g., as part of a reading assignment
from their textbook) and what content they will need to acquire in-class via a combination of lecture and active learning.

I recommend minimizing the amount of class time students spend copying down your derivations of equations.  That does not mean that students should never see a derivation.  Refer your students to the relevant
pages in their textbook where a derivation is shown, or create a video of yourself performing the derivation that students can watch outside of class.  You can hold students responsible for reading these pages or watching
the video by requiring them to answer a few questions that they will submit.  These can be graded either for correctness or for completion, depending on your goals.  This approach can also be used to introduce students
to the declarative knowledge they will need (e.g., the different phases of the interstellar medium), thereby freeing up class time for more active learning, including in-class problem solving scaffolded by PI questions.

As stated earlier, when you choose a quantitative problem for your students to do in class, make sure that the problem will require students to apply as much astrophysical information and as many problem-solving techniques
as possible.  You do not have an unlimited amount of class time with your students, so your goal is to get students to talk to each other and to you about as many of the key decision points in the problem solving process 
as possible.  This will help students develop more expert-like understandings of the course material and problem-solving techniques.

You may still need to devote some class time to lecture, but make sure that lecture is focused on providing students with just enough information so they can start working on a problem.  For example, 
if students have never encountered the method of separation of variables for finding solutions to partial differential equations, then you will need to introduce and explain this method.  But once it has been
introduced and explained, resist the temptation to do several example problems for your students.  Instead, make them use this method to solve one or more problems.

Once you present a problem to students, give them a couple of minutes to work on the problem in collaboration with their neighbors.  Do not immediatelty launch into the PI questions.
Students need time to read and interpret the problem.  They should also have time to begin to decide which pieces of their knowledge they will need to use to figure out the solution.  You probably do 
not want to give students enough time to actually arrive at a solution before asking your first PI question; that may use up a substantial fraction of your class time, and many students may have
become stuck, given up, and disengaged from the class after the first five minutes spent on solving the problem.  I have found that students need two to five minutes to work on the problem before I present the
first PI question.  Two to five minutes is usually enough time for students to read and interpret the problem, formulate their initial ideas of how to solve it, and, most importantly, realize where they may be stuck.
During this time, circulate around the room, listen to students' conversations, and answer questions as they arise.  By listening to your students, you may discover that they are experiencing difficulties
that you did not anticipate and that you can immediately address (and perhaps incorporate into future classes).  You will also get a sense of when many of your students begin to get stuck, which
is when you can start asking your PI questions.  Students will appreciate the need for the PI questions, and the conversations they foster, once they realize that they are stuck.

You should use the best practices for PI when posing your questions to your students and collecting their responses.
Brissenden and Prather have developed a comprehensive guide for implementing PI questions, covering how to pose the question to your students, when to have them discuss their answers with their peers,
how to have students vote, and how to communicate the results of a vote with your students \cite{TPS}.  Their guide contains specific language, developed over many years of classroom experience, that you can use to effectively
communicate to students what you expect them to be doing at each stage during a PI question.  Instructors that are unfamiliar with these best practices should review this guide.

After you have led your class through all of your PI questions, give them some time to complete solving the original problem.  The PI questions will have helped your students unpack many of the key steps
they need to make in order to arrive at the solution.  Give them the chance to get to that solution.  During this time, I have students again work collaboratively with their neighbors.  Once most of the 
class has arrived at a solution, show your solution to the problem, and debrief any lingering issues or questions that your students might have.

\section{Conclusions}
\label{conc}
In this paper I have used the example of the pressure integral to illustrate how many pieces of information and problem-solving decisions are obscured by the all-too brief
solutions that often appear in upper-division textbooks and their associated lectures.  
When an instructor takes the time to list all of the hidden intellectual steps in a calculation, then they can use that list to develop active learning materials for their class.
This paper models how PI questions can be generated from such a list.  This effort is beneficial in two ways.  First, it allows both the instructor and the 
students to better recognize all the pieces of conceptual knowledge and problem-solving abilities that students must acquire.  Second, it enables the instructor to infuse their
class with active learning, which is well-known to elevate student performance and reduce achievement gaps between demographic groups.  Using an active learning 
technique, such as PI, to help 
improve students' problem-solving abilities is a far more effective use of class time than the traditional approach, which consists of students transcribing an instructor's 
solutions and derivations.

The need for models of how to
incorporate active learning into an upper-division course is especially acute in astronomy.  Astronomy education researchers have yet to focus much attention on the 
upper-division, in contrast to the extensive work done on upper-division physics by physics education researchers.  This situation will hopefully change in the coming years.
But in the meantime, there is no need for conscientious instructors of upper-division astronomy courses to stick with ineffective, lecture-only instruction.  They can follow
the model laid out in this paper and the guidelines of Section \ref{dev_imp} to develop PI questions that address both conceptual and quantitative difficulties.  Undertaking such an effort can be rewarding for both the
instructor and their students.  Doing so will also provide a great service to our majors, the future of the discipline, by raising their content mastery and problem-solving skills.

\begin{acknowledgments}
I am indebted to both Edward Prather and Rica French, with whom I have had many stimulating conversations on this topic.  This paper would not have been possible 
without their perspectives and feedback.  I would also like to thank the anonymous referees, whose feedback immeasurably improved the quality of this paper.
\end{acknowledgments}


\begin{thebibliography}{5}

\bibitem{cabanela03} J.\ Cabanela and B.\ Partridge, ``So What IS the Astronomy Major?" \emph{Astron.\ Educ.\ Rev.}, {\bf 2}, 67-84 (2003).

\bibitem{zeilik03} M.\ Zeilik and V.\ J.\ Morris, ``An Examination of Misconceptions in an Astronomy Course for Science, Mathematics, and Engineering Majors,"
\emph{Astron.\ Educ.\ Rev.}, {\bf 2}, 101-119 (2003).

\bibitem{freeman14} S.\ Freeman, S.\ L.\ Eddy, M.\ McDonough, M.\ K.\ Smith, N.\ Okoroafor, H.\ Jordt, and M.\ P.\ Wenderoth, ``Active learning increases student performance 
in science, engineering, and mathematics," \emph{PNAS}, {\bf 111}, 8410-8415 (2014).

\bibitem{black98} P.\ Black and D.\ Wiliam, ``Inside the Black Box: Raising Standards Through Classroom Assessment," \emph{Phi.\ Del.\ Kap.}, {\bf 86}, 139-148 (1998).

\bibitem{eddy14} S.\ L.\ Eddy and K.\ A.\ Hogan, ``Getting Under the Hood: How and for Whom Does Increasing Course Structure Work?" \emph{CBE Life Sci.\ Educ.},
{\bf 13}, 453-468 (2014).

\bibitem{lorenzo06} M.\ Lorenzo, C.\ H.\ Crouch, and E.\ Mazur, ``Reducing the gender gap on the physics classroom," \emph{Am.\ J.\ Phys.}, {\bf 74}, 118-122 (2006).

\bibitem{rudolph10} A.\ L.\ Rudolph, E.\ E.\ Prather, G.\ Brissenden, D.\ Consiglio, and V.\ Gonzaga, ``A national study assessing the teaching and learning of introductory 
astronomy. Part II. The connection between student demographics and learning," \emph{Astron.\ Educ.\ Rev.} {\bf 9}, 010107 (2010).

\bibitem{bailey18} J.\ M.\ Bailey and J.\ D.\ Plummer, ``Editorial: Focused Collection: Astronomy Education Research," \emph{Phys.\ Rev.\ Phys.\ Educ.\ Res.}, {\bf 14},
010004 (2018).

\bibitem{bailey05} J.\ M.\ Bailey and T.\ F.\ Slater, ``Resource letter AER-1: Astronomy education research," \emph{Am.\ J.\ Phys.}, {\bf 73}, 677-685 (2005).

\bibitem{lelliott10} A.\ Lelliott and M.\ Rollnick, ``Big ideas: A review of astronomy education research 1974-2008," \emph{Int.\ J.\ Sci.\ Educ.}, {\bf 32}, 1771-1799 (2010).

\bibitem{loverude15} M.\ E.\ Loverude and B.\ S.\ Ambrose, ``Editorial: Focused Collection: PER in Upper-Division Physics Courses," \emph{Phys.\ Rev.\ ST Phys.\ Educ.\
Res.}, {\bf 11}, 020002 (2015).

\bibitem{mazur97} E.\ Mazur, \emph{Peer Instruction: A User's Manual} (Prentice Hall, Inc., Upper Saddle River, NJ, 1997).

\bibitem{lyman81} F.\ Lyman, ``The Responsive Classroom Discussion," in \emph{Mainstreaming Digest}, ed. A.\ S.\ Anderson (University of Maryland College of
Education, College Park, MD, 1981), pp. 109-113.

\bibitem{dancy10} M.\ Dancy and C.\ Henderson, ``Pedagogical practices and instructional change of physics faculty," \emph{Am.\ J.\ Phys.}, {\bf 78}, 1056-1063 (2010).

\bibitem{manogue06} C.\ A.\ Manogue, K.\ Browne, T.\ Dray, and B.\ Edwards, ``Why is Amp\`{e}re's law so hard? A look at middle-division physics," \emph{Am.\ J. Phys.},
{\bf 74}. 344-350 (2006).

\bibitem{wilcox13} B.\ R.\ Wilcox, M.\ D.\ Caballero, D.\ A.\ Rehn, and S.\ J.\ Pollock, ``Analytic framework for students' use of mathematics in upper-division physics," 
\emph{Phys.\ Rev.\ ST Phys.\ Educ.\ Res.}, {\bf 9}, 020119 (2013).

\bibitem{dray99} T.\ Dray and C.\ A.\ Manogue, ``The vector calculus gap: Mathematics $\neq$ physics," \emph{Primus}, {\bf 9}, 21-28 (1999).

\bibitem{redish05} E.\ F.\ Redish, ``Problem solving and the use of math in physics courses," arXiv:physics/0608268v1.

\bibitem{carroll07} B.\ W.\ Carroll and D.\ A.\ Ostlie, \emph{An Introduction to Modern Astrophysics}, $2^{\textnormal{nd}}$ ed.\ (Pearson Education, Inc., San Francisco, CA, 2007), p.\ 289.

\bibitem{harwit06} M. Harwit, \emph{Astrophysical Concepts}, $4^{\textnormal{th}}$ ed.\ (Springer, New York, NY, 2006), p.\ 115.

\bibitem{maoz07} D.\ Maoz, \emph{Astrophysics in a Nutshell} (Princeton University Press, Princeton, NJ, 2007), p.\ 44.


\bibitem{scher19} B. P. Schermerhorn and J. R. Thompson, ``Physics students' construction and checking of differential volume elements in an unconventional coordinate
system," \emph{Phys.\ Rev.\ Phys.\ Educ.\ Res.}, {\bf 15}, 010112 (2019).

\bibitem{TPS} G.\ Brissenden and E.\ Prather, ``Think-Pair-Share: A Revised `How-To' Guide," 
https://www.aapt.org/Conferences/newfaculty/upload/ CAE\_TPS\_How\_To.pdf

\bibitem{prather09} E.\ E.\ Prather, A.\ L.\ Rudolph, G.\ Brissenden, and W.\ M.\ Schlingman, ``A national study assessing the teaching and learning of introductory astronomy. 
Part I. The effect of interactive instruction," \emph{Am.\ J.\ Phys.}, {\bf 77}, 320-330 (2009).

\end{thebibliography}
\end{document}